# Human-centered transparency of grasping via a robot-assisted minimally invasive surgery system

Milstein Amit, *Student Member IEEE*, Ganel Tzvi, Berman Sigal, *Member IEEE*, and Nisky Ilana, *Member IEEE*

*Abstract*— We investigate grasping of rigid objects in unilateral robot-assisted minimally invasive surgery (RAMIS) in this paper. We define a human-centered transparency that quantifies natural action and perception in RAMIS. We demonstrate this human-centered transparency analysis for different values of gripper scaling – the scaling between the grasp aperture of the surgeon-side manipulator and the aperture of the surgical instrument grasper. Thirty-one participants performed teleoperated grasping and perceptual assessment of rigid objects in one of three gripper scaling conditions (*fine*, *normal*, and *quick*, trading off precision and responsiveness). Psychophysical analysis of the variability of maximal grasping aperture during prehension and of the reported size of the object revealed that in *normal and quick* (but not in the *fine*) gripper scaling conditions, teleoperated grasping with our system was similar to natural grasping, and therefore, human-centered transparent. We anticipate that using motor control and psychophysics for human-centered optimizing of teleoperation control will eventually improve the usability of RAMIS.

*Index Terms*—Grasping, Human-robot physical interaction, Sensorimotor control, Robot-assisted surgery, Telerobotics

## I. Introduction

During the course of our life, we interact with and perceive our world through our hands and eyes. In robot-assisted minimally-invasive surgery (RAMIS), a surgeon utilizes local manipulators to teleoperate remote surgical instruments inside the body of a patient while observing the surgical environment through a three-dimensional endoscopic camera. Hence, the surgical tools and the endoscopic camera serve as the remote hands and eyes of the surgeon to act upon and perceive the surgical environment. In RAMIS, the patient benefits from all the advantages of standard minimally-invasive surgery including less recovery time, blood loss, and pain [1]. In addition, the surgeon gains improved tool manipulation due to additional degrees-of-freedom, motion scaling, improved precision, and better vision of the surgical site [2]. These have contributed to a wide adoption of RAMIS in many surgical procedures [3]. However, the evidence about the improvement of patient outcome in RAMIS when compared to standard minimally-invasive surgery is mixed in certain procedures, and the adoption of RAMIS is still limited in others [3]. We suggest that some of the current limitations in RAMIS could be mitigated by optimizing the system's controller such that the surgeon's action and perception will be natural and similar to open surgery.

Generally, teleoperation controllers synchronize between the motions of the local and remote manipulators; bilateral force reflecting teleoperation also presents the forces that are applied by the environment to the user. The fidelity of teleoperation is defined as its transparency. Ideally, in a transparent system, the operator's intentions are executed accurately, thus the accurate perception of the environment [4]. Traditionally, transparency is defined only for bilateral teleoperation. Different ways to measure transparency were proposed, including comparison of motions and forces of the local and remote manipulators [5], or the impedances transmitted via the teleoperation channel [6]. However, in state-of-the-art RAMIS systems, the surgeons do not receive force feedback [7]. Moreover, the classical measures focus on the system rather than the action and perception of the surgeon. Therefore, to analyze and improve RAMIS control, a measure of transparency suitable for unilateral teleoperation without force feedback is necessary.

Evaluating the relation between the transparency of teleoperation systems and task performance is important. For example, RAMIS systems allow better visualization, tremor filtering and changing the scaling between the movements of the surgeon and the movement of the surgical tool. Scaling can be lowered to improve precision, and it can be increased to have a faster and more responsive operation. Such performance-enhancing design is not transparent according to the classical definition, where both sides of the teleporter have identical kinematics and dynamics.

Recently, a human-centered transparency approach was proposed [4]. This approach is based on several examples of gaps between the action and perception in human sensorimotor control [8]–[10]. In this study, we adopt this human-centered approach and define transparency in RAMIS from the perspective of the surgeon: (1) her actions are natural and similar to the actions during open surgery, and (2) her perception is directly interacting with the patient. Using this approach, we identify critical conditions that break human-centered transparency.

This research was supported in part by the Helmsley Charitable Trust through the Agricultural, Biological and Cognitive Robotics Initiative and by the Marcus Endowment Fund both at Ben-Gurion University of the Negev, and by the Israeli Science Foundation (grant number 823/15). AM is supported by a fellowship from the Ministry of Science and Technology, Israel.

A. Milstein and I. Nisky are with Department of Biomedical Engineering at Ben-Gurion University of the Negev, Beer-Sheva, Israel.

S. Berman is with the Department of Industrial Engineering and Management at Ben-Gurion University of the Negev, Beer-Sheva, Israel

T. Ganel is with the Department of Psychology at Ben-Gurion University of the Negev, Beer-Sheva, Israel.

(Correspondence e-mail: nisky@bgu.ac.il).

Here, we focus on the human-centered transparency of grasping in RAMIS. Grasping and manipulating rigid and soft objects, such as needles and tissue, is a crucial part of the majority of RAMIS procedures. Many RAMIS instruments have a grasper as their end-effector, and the surgeon uses the gripper of the local manipulator to teleoperate these graspers using their thumb and finger. An example that highlights the importance of grasping in RAMIS is the incorporation of the peg transfer task in the Fundamentals of Robotic Surgery [11].

In this paper, we investigate how the gripper scaling – the ratio between the surgeon side manipulator's gripper aperture angle to that of the surgical tool – affects the action and perception of the user. In commercial systems, scaling is only available for the translation of the tool, and can only affect manipulation of objects. Here, we focus on grasping of objects, and therefore, chose to focus on the gripper scaling. This scaling is important in unilateral telesurgery as it allows adjusting the sensitivity of the tool's graspers for grasping differently sized tissue and blood vessels, needles and accessories in a natural manner. Specifically, we focus on the reach-to-grasp, and therefore, to avoid the complications of interaction with a soft tissue, we chose the simpler task of grasping rigid objects.

Natural grasping has been studied extensively in the fields of human motor control and psychophysics. When grasping an object, we first reach towards the object and grasp it only when the hand is close enough [12]. During the hand transport, we open our fingers to a maximum grip aperture (MGA) that is larger than the object, but is not necessarily equal to the maximum capacity of our fingers. Importantly, this aperture is proportional to the size of the object, and allows our hand to stably grasp the object perpendicular to its surface. However, it is not established whether the kinematics of grasping in teleoperation and RAMIS is similar to natural grasping.

The perception of the size of objects is consistent with Weber's Law [13], stating that the discrimination sensitivity of the size (the just noticeable difference, JND) is proportional to the size of the object [14]. In contrast, the variability of the maximum grip aperture during the transport does not depend on object size, and violates Weber's Law [14]. These findings demonstrate a dissociation between action and perception in natural grasping. Importantly, this dissociation exists even when the perception is reported using manual estimation (or pantomimed grasping) rather than verbal or forced choice reports [15], [16], highlighting different neural processing between action and perception rather than differences in the output modality. In our study, we investigate the effect of a RAMIS setup, in which the remote environment is accessible only through a proxy tool and a camera, on the dissociation between action and perception of the human operator.

Specifically, we focus on grasping objects of various diameters, and the perception of their size. We define the dissociation between action and perception in the remote's grip aperture variability as an indication for natural interaction. In a human-centered transparent system, which induces natural grasping, we expect that, first, the kinematics of grasping is characterized by a maximum grasping aperture that occurs during the reach-to-grasp motion that is proportional to the size of the object. The variability of this peak grip aperture violates Weber's law and does not depend on the size of the grasped object. Second, the variability of pantomimed perceptual assessments obeys Weber's law and linearly increases with the diameter of the object. This is important because such a dissociation is a necessary condition for asserting that similar underlying neural control mechanism mediate sensorimotor control in a teleoperation setup, e.g. RAMIS.

To demonstrate this transparency analysis, we use a psychophysical experiment to compare the action and perception in a teleoperated RAMIS setup and investigate whether under different gripper scaling conditions such as dissociation exists. We found that as long as the relation between the gripper scaling, the size of the gripper, and the size of the objects did not limit the possible grasping apertures, our unilateral teleoperated RAMIS system allowed a natural teleoperated grasping with a dissociation between action and perception, and was human-centered transparent. However, in our simple grasping task, we did not establish a link between transparency and performance. We also highlight that our transparency condition may be insufficient, and it should be used in conjunction with a performance optimization.

## II. METHODS

### A. Hardware

Our teleoperated RAMIS setup (Fig. 1A) consists of two parts: the surgeon's operating console (local operator), and the surgical robot (remote operator). The surgeon's operating console, depicted in Fig. 1B, consists of a metal frame, a manipulator and a 3D vision system, all connected to a computer with an Intel Core Xeon E5-1620 v3 processor. The local manipulator is a SIGMA 7 (Force Dimension) haptic device, which has seven degrees-of-freedom and a built-in grasper, which has a 30 degrees aperture range (approximately 3 cm range in grip aperture of the user). The vision system consists of two Flea3.0 (PointGrey) USB cameras equipped with 16 mm f1.8 compact instrumentation lenses (Edmund Optics), and a 3D viewer, HMZ-T3W (Sony), fixed to the surgeon's console frame. The experimental scene is acquired by the cameras and presented to the operator at 60 Hz and a resolution of 1080p for each eye. This presents the operator with a 3D view of the environment. The surgical robot was the Raven II (Applied Dexterity) [17] (Fig. 1C). Each of the two arms of the Raven II is a cable-driven seven degrees-of-freedom manipulator with a gripper end effector. Each arm was connected to a separate controller, identical to the one described in [17], and through USB cables to a computer with an Intel Core Xeon E5-2603 v3 processor. In this study, we only use the right arm of the Raven II. The communication between the local and remote operators was established over the University's local area network (LAN), using a UDP/IP socket.

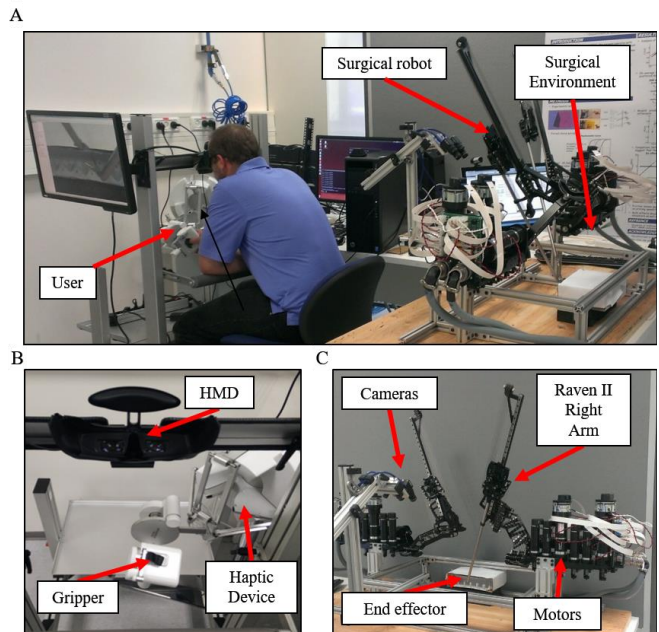

Fig. 1. Experimental setup: (A) an overview of the setup: the local user, on the left hand side of the picture, teleoperating the surgical robot in the surgical environment, on the right hand side; (B) local side console consists of the head-mounted device (3D Viewer), fixed to user interface frame, which displays subjects with the surgical scene and the haptic device with the gripper handle; (C) remote side consists of the RAVEN II Surgical Research System and the two high definition cameras.

*B. Software, Control and Communication*

We implemented a unilateral position control scheme (Fig. 2) to teleoperate the surgical robot using the local manipulator. Our architecture, kinematics and control were based on the native RAVEN II controller [17], [18] with the changes specified below. The user's state vector ($X_{user}$) is recorded by the local manipulator. This state vector consists of the Cartesian position, orientation, and grip aperture angle ($X_{user} = [x, y, z, \alpha_x, \alpha_y, \alpha_z, \alpha_{gripper}]$) of the local manipulator. During teleoperation, no forces other than the passive dynamics of the Sigma7 device ($F_{dynamic}$) were applied on the user via the software.

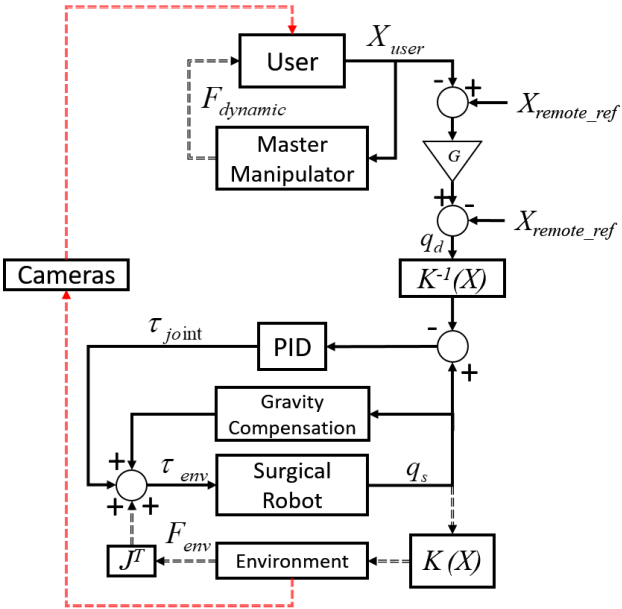

Fig 2. The teleoperation control system, implementing position control architecture. On the upper side is the user operator, and on the bottom is the surgical environment. The visual data acquired by the cameras is the red dashed line. The forces applied by the dynamics of the haptic manipulator on the user and the environmental forces acting on the surgical manipulator are a dashed black line.

To align the local and remote workspaces after instrument clutching, reference positions were stored for the local and remote manipulators ($x_{local\_ref}$ and $x_{remote\_ref}$, respectively). The desired pose of the remote manipulator was:

$$x_{remote\_desired,i} = x_{remote\_ref,i} + g_i \cdot \left(x_{user,i} - x_{local\_ref,i}\right), \quad (1)$$

where $x_{user,i}$ is the i$^{th}$ variable of the user's state vector and $g_i$ is the scaling of the i$^{th}$ state variable. The values of the Cartesian, and orientation scaling were $g_{Cartesian}=0.5$ for all Cartesian variables and $g_{Orientation}=0$ for all orientation variables; both were held constant for all groups and all subjects. The gripper scaling, which determines the scaling between the local and remote opening angles ($g_{gripper}$), was set to $g_{gripper}=3$ (*fine*), 5 (*normal*), or 7 (*quick*), and was held constant for each subject throughout the entire experiment. The *fine* gripper scaling is smaller than the *normal* gripper scaling, meaning a larger grip aperture opening by the user is needed to achieve the same aperture as in the *normal* gripper scaling (Fig. 3A). The *quick* gripper scaling is larger than the *normal* gripper scaling, meaning a smaller grip aperture opening by the user is needed to achieve the same aperture as in the *normal* gripper scaling.

The desired trajectory ($x_{remote\_desired}$) was transmitted over the communication channel to the software of the remote manipulator, where it was transformed into the desired joint angles $q_{desired}$ and then into the desired motor angles:

$$q_{desired} = K^{-1}\left(X_{remote\_desired}\right), \quad (2)$$

where $K^{-1}(X_{remote\_desired})$ is the inverse kinematics of the Raven II. The desired angles were then transferred to a PD controller:

$$\tau_{PD,i} = k_{p,i}\left(q_{desired,i} - q_{s,i}\right) + k_{d,i}\left(-\dot{q}_{s,i}\right) \quad (3)$$

where $\tau_{PD,i}$ is the i$^{th}$ joint torque, $q_{s,i}$ is the current i$^{th}$ joint state – estimated using the Raven native mapping from encoder readings to joint angles – and $\dot{q}_{s,i}$ is the joint i$^{th}$ velocity, which was estimated using a back differentiation from previous motor readings. The gains for the proportional, $k_p=[0.3, 0.3, 0.15, 0.009, 0.05, 0.02, 0.02]$, and derivative, $k_d=[0.008, 0.008, 0.01, 0.001, 0, 0, 0]$, terms were chosen empirically for smooth and stable operation. To reduce potential experimental confounding factor associated with orientation control, we kept the orientation of the remote teleoperated tool constant. Finally, feedforward gravity compensation torques ($\tau_{GC}$) [19] were added, and the resulting torques ($\tau_{joint}$) were transformed to motor commands, and applied to the hardware controllers.

*C. Participants and experimental conditions*

Thirty-one right-hand-dominant participants (14 females, 25.5±2.4) took part of our study. All participants were students at Ben-Gurion University of the Negev, who signed a written consent, and were compensated for their participation regardless of their results and experiment completion. They were randomly assigned for one of three gripper scaling groups without balancing of age, gender, or any other category. Each group had a different gripper scaling: (1) *fine*, where $g_{gripper}=3$ (N=10, 5 females); (2) *normal*, where $g_{gripper}=5$ (N=11, 5 females); and (3) *quick*, where $g_{gripper}=7$ (N=10, 4 females). Gripper scaling conditions were chosen to allow successful task completion as a demonstration of our approach. The *normal* scaling was chosen empirically to allow for comfortable grasping of the objects with the tool's gripper without approaching any physical limits at the local manipulator. The *quick* scaling was chosen to allow faster reaction of the remote tool, and the *fine* scaling was chosen to allow better accuracy. The gripper scaling was constant for each subject and did not change throughout the experiment nor between the experiments. Our novel analysis of human-centered transparency relies on evaluating how natural the teleoperated grasping is by comparing the variability of grasping of several differently-sized objects to the variability of the perceptual assessments of their size. Therefore, all participants performed two back-to-back experiments: (1) action and (2) perception, both with the same gripper scaling. The order of the experiments was balanced across participants within each gripper scaling groups.

*D. Protocol*

Participants sat in front of the local console, viewed the instructions and remote environment via the 3D viewer, and held the gripper of the haptic device in their right hand, and a computer keyboard was placed on their lap to be used with their left hand. In each experiment, participants grasped five cylindrical objects, differing in diameter (4mm, 6mm, 8mm, 10mm, 12mm), as depicted in Fig. 3B. Each experiment consisted of 110 trials, and included 22 grasps of each of the five objects. The order of objects was pseudo-randomized and predetermined for each experiment, and did not change across participants, such that in every block of 10 trials, participants grasped each object twice. The first ten trials were training and were not used in data analysis.

In the action experiment, participants were instructed to perform a remote reach-and-grasp task, using the teleoperation system. Written instructions were embedded in the visual display at the relevant stage of the task. Participants started in a closed grasp – the grasper of the haptic device was fully closed, but their finger and thumb did not, in fact, touch each other - viewing a black screen, with a displayed message "*get ready*". An object was placed 40 mm in front of the closed gripper. After 300 milliseconds, a message "*GO*" was displayed, and participants reached, grasped, lifted, and released the object. Upon releasing the object, participants hit the spacebar key on the keyboard to indicate they had finished. Once a trial was finished, the participants' hand was guided by the haptic device back to the starting position for the next trial.

In the perception experiment, upon the initial display of the remote environment, participants were instructed to first show their estimation of the object's size using a pantomimed grasp gesture with the remote side instrument (similarly to [14]), and a "*Show Grasp and press Space*" message was displayed on the visual display. To make sure that in the perception experiment, the participants received identical information about the objects and the teleoperation system as in the action experiment, we asked the participants to complete the lifting movement [14]. Upon finishing the perceptual part of the task, participants were instructed to hit the spacebar key, and, following a "*GO*" message, proceed with the reach, grasp, lift, and release sequence. Then they had to once again hit the spacebar, to indicate trial completion. Once a trial was finished, the participants' hand was guided by the haptic device back to the starting position for the next trial. The data from the reach-to-grasp of the perception experiment was not analyzed.

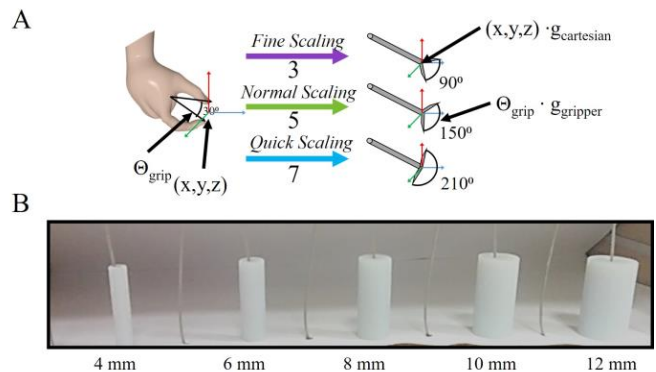

Fig. 3. Experimental conditions and objects: (A) Illustration of gripper scaling between local (hand) and remote (tool) sides in teleoperation. (B) The five cylindrical objects that the participants lifted in our experiment.

*E. Data analysis*

We recorded the remote tool trajectories and gripper aperture from the native state estimator of the Raven II at 1 KHz. We then down-sampled to 100 Hz, and low pass-filtered at 10 Hz using a 4th order zero-lag Butterworth filter using the *filtfilt()* function in MATLAB® for further analysis. This resulted in an 8th order filter with a 9 Hz cutoff frequency and no phase-shift. Examples of a filtered path and the trajectories of the endpoint and the aperture of the gripper are depicted in Fig. 4 and Fig. 5, respectively. We analyzed three aspects of our task: (1) action and perception metrics, (2) timing metrics, and (3) transport and grasp kinematics metrics. To quantify these, we defined the following metrics:

*1) Analysis of natural action and perception*

To test the effect of the gripper scaling on how natural grasping was in the action and perception experiments, we extracted the following metrics:

**Maximum grip aperture (MGA).** The maximal grip aperture angle of the remote side gripper in each trial in the action experiment was measured in radians during the reaching phase (green rectangle in Fig. 5B). We calculated the mean and standard deviation of MGA per object size for each participant. If our system induces a natural action, the mean MGA is expected to be proportional to the size of the object [9], [20], and the standard deviation of MGA is not expected to depend on object size, in violation of Weber's law [14].

**Pantomimed object size (PS).** The grip aperture of the remote side instrument in the pantomimed grasp gesture phase of the perception task is measured in radians. We calculated the mean and standard deviation of PS for each subject and object size. If our system induces a natural perception, the mean PS is expected to be proportional to the size of the object, and the standard deviation of the PS is expected to increase with the size of the object, in accordance with Weber's Law.

*2) Analysis of timing*

We assume that difficult tasks take longer to plan and

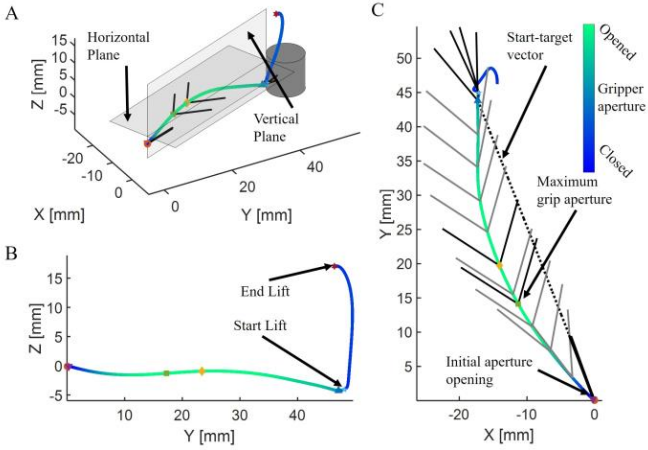

Fig. 4. An example of a path of a single reach-to-grasp and lift trial at the remote (slave) side. (A) 3D view of the task. The blue-green line represents the movement as projected on (B) X-Y (horizontal – "top view") plane and (C) the start-target plane ("side view"), noted by the dotted line in (A). The color of the line corresponds to the gripper aperture during that segment of the movement (color-bar), where blue represents fully closed and green represent fully opened. The black lines in (A,B) represent the grasper's posture at major points of interest.

execute. Therefore, to test whether gripper scaling had an effect on the task difficulty, we calculated these timing metrics:

**Reaction time.** The duration between object appearance and transport onset – marked with a red circle marker in Fig. 5A. The longer this duration is, the more complex the task is considered since it requires more pre-programming [10].

**Transport time.** The duration between moving from the start position to the target – shaded pink area in Fig. 5A, between the red circle and blue upward pointing triangle.

**Total time.** The total duration of the task, from object display to the end of lifting, marked as the red hexagram in Fig. 5.

**Perception time.** The time duration between object appearance and the time it took the participant to show the PS in the perception experiment.

*3) Analysis of task kinematics*

In the action experiment, we used the following metrics for the kinematics of the transport and the grasping:

**Transport path length.** The total distance traveled by the gripper endpoint from object appearance to the end of transport, measured in millimeters (mm), shaded pink area in Fig. 5. This was calculated by numerical integration of the endpoint trajectory data points from the start position to the end of the transport phase. A straight reach would result in the shortest path of 40 mm. Curved reaches would yield longer distances.

**MGA timing fraction.** The timing of MGA normalized by transport time. This was the time stamp of the MGA, divided by total reach time. This metric can also have negative values, e.g. when a participant opens their aperture to a maximum before transport onset. However, in our analysis, we set negative values to zero indicating that at the beginning of the transport, MGA was already set. In natural reach-to-grasp movements, this metric value is approximately 0.6 to 0.7 [12].

**Peak endpoint speed.** The peak speed during the transport– marked as yellow diamond in Fig. 5A.

**Peak grip aperture speed.** The peak speed of the grip aperture during the transport– can be seen in Fig. 5B.

*F. Statistical analysis*

For each of the above metrics, we calculated the mean across different lifts for each participant and object. For MGA and PS, we also calculated the standard deviation across different lifts for each participant and object. For each of these 11 statistics as a dependent variable, we fitted a generalized mixed model. The independent factors were gripper scaling (categorical, three levels – *fine*, *normal*, and *quick*, fixed between-subjects effect), object diameter (continuous, fixed within-subject effect), their interaction, and the participant (categorical, random effect). From the model, we extracted the slope coefficient for each gripper scaling, and the adjusted mean, calculated at the center of the object diameter range (8 mm). We examined the QQ plots to verify the assumption of normality of the residuals, and used Bartlett's test for homogeneity of variances. The only metrics that violated normality were timing related, hence, we log-transformed the data prior to fitting of the model. We also performed preplanned comparisons to test for statistical significance differences between gripper scaling groups for the above metrics. All data and statistical analyses were performed using custom-written MATLAB code.

## III. RESULTS

A typical grasping path is depicted in Fig. 4, and the corresponding endpoint and gripper aperture trajectories are depicted in Fig. 5. The object was displayed to the participant, and after the initial response time (the gray-shaded areas in Fig. 5), the participant reached to the object, grasped and lifted it to approximately 20 mm in height, and finally released it from that height. Fig. 4A depicts the 3D path, and Fig. 4B and Fig. 4C depict the X-Y (horizontal – "top view") plane and the start-target plane ("side view"), respectively. Seventeen participants (5 in the *quick*, 5 in the *normal*, and 7 in the *fine groups*) had similar paths and trajectories to those depicted in Fig. 4 and Fig. 5 - they started the reaching movement and soon after that started to open their grip aperture until reaching the MGA (green square). Ten of the participants (4 quick, 4 normal, 2 fine) opened the gripper before starting the movement, and four participants (1 quick, 2 normal, 1 fine) were very late in the movement, which is atypical in real-world natural grasping movements. All of the participants were included in all performed analysis.

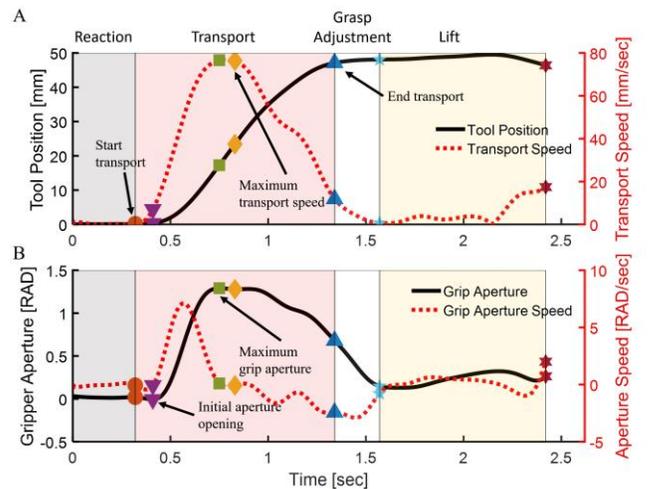

Fig. 5. An example of the trajectories of the movement and the grip aperture of a single reach-to-grasp and lift trial. (A) Position (solid black line), projected on the start-target vector (denoted as the dotted line in Fig. 4), and speed of the projected movement (dotted red line); (B) grip aperture of the graspers (solid black line), and speed of the grip aperture (dotted red line).

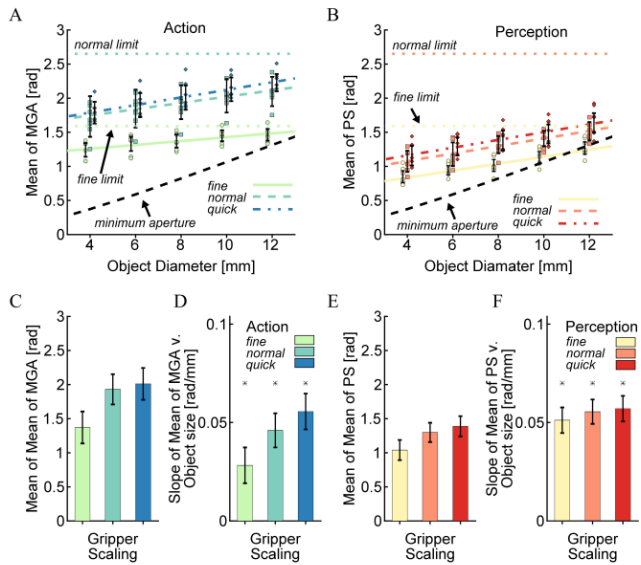

Fig. 6. Mean of the maximum grip aperture (MGA) in the action experiment (A, C, D) and mean of the perceptual assessment of object size (PS) in the perception experiment (B, E, F). (A, B) grip aperture of the remote side instruments as a function of grasped object diameter. The small colored symbols are individual subjects' mean. The black large symbols and the error bars are the means across subjects and their 95% CI, for each gripper scaling. The colored lines are the regression that was fit to the subjects' mean maximum grip apertures. The black dashed lines represent the minimal gripper aperture needed to grasp an object. The dotted horizontal lines indicate the maximum possible grip aperture for each scale (resulting from a limit on the opening aperture of the master manipulator). Larger values were measured in 4.4% of all action trials and 0.2% of all perception trials due to hardware limitations. The maximum possible grip for the *quick* scale is outside of the range of the figure. (C, E) The estimated adjusted mean aperture (calculated at the 8mm object) for each gripper scaling. (D, F) The slope of the regression model of mean MGA as a function of object size for each gripper scaling. The error bars are 95% confidence interval for the estimated mean, asterisks indicate the statistical significance of the slope coefficient being different than zero (* p<0.05 and ** p<0.01).

In the remainder of this section, we demonstrate our human-centered transparency assessment, and report the effect of gripper scaling on grasping in teleoperation. Then, to evaluate whether the gripper scaling affected task difficulty and complexity, we analyze its effect of on timing, and finally, we compare several kinematics metrics between the different

TABLE I
STATISTICAL ANALYSIS OF TRANSPARENCY METRICS*

|  | Gripper Scaling | Object size | Interaction |
|---|---|---|---|
| Mean of MGA | **$F(2,33.99)=8.00$** **p=0.001** | **$F(1,121.00)=539.22$** **p=0.000** | **$F(2,121.00)=18.06$** **p=0.000** |
| STD of MGA | $F(2,141.07)=0.09$ p=0.912 | $F(1,121.00)=0.01$ p=0.932 | **$F(2,121.00)=3.62$** **p=0.030** |
| Mean of PS | **$F(2,43.78)=7.09$** **p=0.002** | **$F(1,121.00)=822.81$** **p=0.000** | $F(2,121.00)=0.84$ p=0.433 |
| STD of PS | $F(2,96.60)=2.29$ p=0.107 | **$F(1,121.00)=14.23$** **p=0.000** | **$F(2,121.00)=8.10$** **p=0.001** |

*Bolded cells are statistically significant (p<0.05).

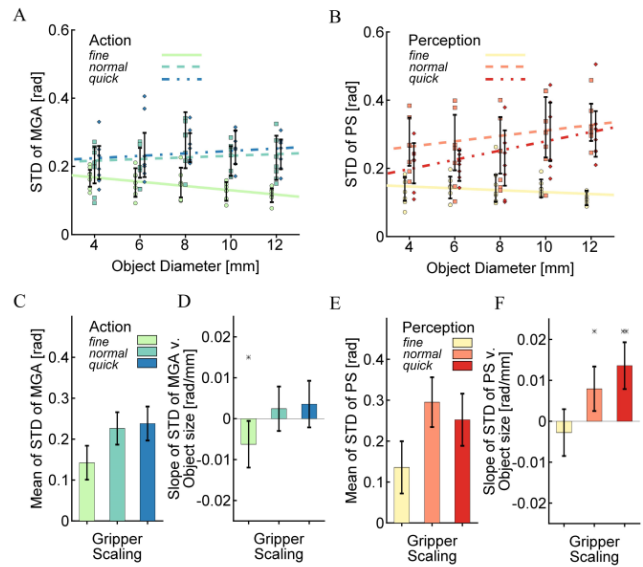

Fig. 7. Standard deviation (STD) of the maximum grip aperture (MGA) of the remote side instrument in action (A, C, D) and of the perceptual assessment of object size (PS), in perception (B, E, F). Line styles are the same as in Fig. 6. (A, B) STD of the grip aperture as a function of grasped object size for each gripper scaling, with a fitted regression line for each gripper scaling. (C, E) The estimated adjusted mean STD (calculated at 8mm) of the grip aperture for each gripper scaling. (D, F) the slope coefficient of the fitted regression line for each gripper scaling.

gripper scaling conditions. Throughout this section, we use the term *significant* to describe effects that are statistically significant at the 0.05 threshold after the necessary corrections for multiple comparisons as needed.

*A. Analysis of natural action and perception*

Table I contains the statistical analysis summary of the mean and standard deviations of the MGA (action) and PS (perception) as a function of gripper scaling, object size, and their interaction in a generalized mixed model. The first two rows contain the F-value and p-value for the main effects of gripper scaling and object size as well as the interaction effect on the mean and standard deviation of the MGA in the action experiment. The last two rows contain the same information for the PS in the perception experiment. Bartlett's test resulted in significant violation of homogeneity assumption between the different scaling groups but not between the objects. Our interest is in the analysis of the dependence of MGA and PS on the size of the object within each group, and therefore, we continue with the analysis, but do not compare contrasts between scaling groups. We also verified that separate regression models for each group yield similar conclusions.

*1) Mean of Maximum Grip Aperture and of Pantomimed Object Size*

Fig. 6 depicts the dependency of the mean maximum grip aperture at the patient-side on the size of the object for the different gripper scaling levels from the action and perception experiments. Fig. 6A and Fig. 6B illustrate the regression lines that were fitted as part of the generalized mixed model to the grip aperture as a function of the object size for each gripper scaling group in the action and perception experiments, respectively. Fig. 6D and Fig. 6F show the slope coefficient for each of the gripper scaling groups, in the action and perception experiments, respectively. All slopes were significantly different from zero – this means that there was a significant dependency of object size for both MGA (action: t=8.56 and p<0.0001, t=14.64 and p<0.0001, t=16.86 and p<0.0001 for the *fine*, *normal* and *quick* groups, respectively) and PS (perception experiment: t=15.28 and p<0.0001, t=14.40 and p<0.0001, t=17.05 and p<0.0001 for the *fine*, *normal* and *quick* groups, respectively) in all gripper scaling groups. This suggests that similar to natural grasping, participants matched their MGA to the size of the object during reaching and their PS during the pantomimed assessment of the size of the object.

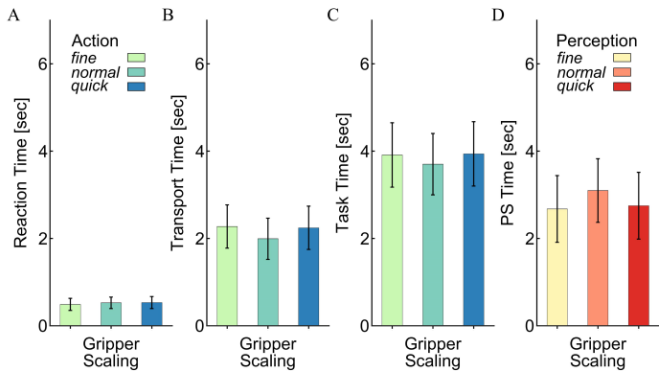

Fig. 8. Analysis of complexity as measured using timing of action (A-C) and perception experiments (D). Color code as in Fig. 6. (A) Reaction time; (B) transport time; (C) total task time, (D) time to perception.

In both experiments, there was also a significant main effect of the gripper scaling on MGA ($F_{2,33.99}$=8.00, p<0.01) and PS ($F_{2,43.78}$=7.09, p<0.01). In the action experiment, there was a significant interaction effect between the scaling and object size ($F_{2,121}$=18.06, p<0.0001).

*2) Standard deviation of Maximum Grip Aperture and of Pantomimed Object Size*

Fig. 7 depicts the dependency of the standard deviation of the grip aperture at the patient-side on the size of the object for the different gripper scaling values from the action and perception experiments. Fig. 7A and Fig. 7B depict the regression lines that were fitted as part of the generalized mixed model to the standard deviation of the grip aperture as a function of the object size for each gripper scaling group in the action and perception experiments, respectively. Fig. 7D and Fig. 7F show the slope coefficient for each of the gripper scaling groups, in the action and perception experiments, respectively.

In the action experiment, the only significant effect was the interaction between the gripper scaling and object size ($F_{2,121}$=3.62, p<0.05). For the *normal* and *quick* gripper scaling groups, the slope, i.e. the dependency between the STD of MGA and object size, was not significantly different from zero (t=0.87, p=0.19 and t=1.22, p=0.11). This means that the variability of the MGA does not depend on the size of the object, for the *normal* and *quick* gripper scaling, and therefore, consistent with natural grasping, the participants violated Weber's law in the control of action via teleoperation. However, interestingly, for the *fine* gripper scaling group, the slope was significantly less than zero (Fig. 7D, t=2.16, p<0.05), which means that the variability of the MGA decreased with the object size. Overall, the standard deviation of MGA appears to be smaller for the *fine* gripper scaling group compared to the *normal* and *quick* gripper scaling (Fig. 7C), but this effect was not significant. In the perception experiment, there was a significant main effect of the object size ($F_{1,121}$=14.23, p<0.0001) and a significant interaction effect between the gripper scaling and object size ($F_{2,121}$=8.10, p<0.01). The slope coefficients were consistent with Weber's law and were significantly larger than zero only for the *normal* and *quick* gripper scaling groups (Fig. 7F, t=2.86, p<0.01 and t=4.65, p<0.0001). Surprisingly, in the *fine* gripper scaling group, the slope coefficient was not significantly different from zero (t=0.94, p=0.17, Fig. 7F).

Taking together the results of the analysis of the mean and standard deviation of the MGA (action) and PS (perception), we conclude that in our RAMIS system in the *normal* and *quick* conditions, teleoperated grasping was similar to natural grasping. In addition, in the *fine* gripper scaling, participants adopted a different, un-natural behavior in both experiments.

*B. Timing*

In addition to examining whether our RAMIS system induces natural action and perception, we also wanted to assess how difficult it was to execute the task using an objective metric. We assume that difficult tasks take longer to plan and execute. Therefore, we examined how the different gripper scaling affected transport times. Fig. 8 shows the different timing metric values for each gripper scaling group.

There was no significant main effect of gripper scaling nor significant interaction effect between gripper scaling and object size on neither the mean reaction time (gripper scaling: $F_{2,43.29}$=0.22, p=0.804, interaction: $F_{2,121}$=0.04, p=0.958 – Fig. 8A), transport time (gripper scaling: $F_{2,35.74}$=0.74, p=0.484, interaction: $F_{2,121}$=0.12, p=0.889 – Fig. 8B), nor total task time (gripper scaling: $F_{2,43.29}$=0.09, p=0.915, interaction: $F_{2,121}$=1.28, p=0.282 – Fig. 8C) in the action experiment. This means that the gripper scaling does not affect the transport, and suggests that the transport and grip aperture movements were planned and controlled separately. There was a significant main effect of object size only on the mean reaction time ($F_{1,121}$=12.45, p<0.01) – the reaction time increased with object size. This suggests that participants considered larger objects more difficult to grasp, which is also found in the studies of natural control of grasping. In the perception experiment, there was no significant effect of any of the factors on the log-transformed perception time (gripper scaling: $F_{2,50.89}$=0.67, p=0.515, object size: $F_{1,121}$=0.25, p=0.615, interaction: $F_{2,121}$=1.16, p=0.318 – Fig. 8D).

*C. Task Kinematics*

Fig. 9 shows the different metrics for each gripper scaling.
*1) Transport Kinematics*

The path length and peak endpoint speed for each gripper scaling group is shown in Fig. 9A and 9C, respectively. There were no significant main effects nor interaction effect on either the peak endpoint speed (gripper scaling: $F_{2,31.01}$=3.06, p=0.061, object size: $F_{1,121}$=3.59, p=0.061 interaction: $F_{2,121}$=1.08, p=0.344) or path length (gripper scaling: $F_{2,32.61}$=0.77, p=0.471,

object size: $F_{1,121}=1.38$, $p=0.243$ interaction: $F_{2,121}=0.84$, $p=0.433$). This further supports our assertion that participants separated the control of transport from the control of grip aperture, hence the gripper scaling did not affect the kinematics of the transport.

*2) Gripper Kinematics*

There were significant main effects of gripper scaling and object size on the peak grip aperture speed (gripper scaling: $F_{2,31}=4.45$, $p<0.05$, object size: $F_{1,121}=146.68$, $p<0.0001$ interaction: $F_{2,121}=2.31$, $p=0.103$). Multiple comparison analysis revealed that the peak grip aperture speed for the *normal* gripper scaling was significantly larger than that of the *fine* gripper scaling group ($t=2.91$, $p<0.005$, Fig. 9D). In contrast, there was no significant main effect on the MGA timing (gripper scaling: $F_{2,40.76}=0.58$, $p=0.565$, object size: $F_{1,121}=2.23$, $p=0.138$, interaction: $F_{2,121}=0.84$, $p=0.433$ – Fig. 9C). However, the grand mean, of 0.53 is slightly lower than the reported values for natural grasping (0.6 to 0.7) [21].

## IV. DISCUSSION

In this paper, we explored a novel approach to examine whether grasping with a unilateral teleoperated Raven II RAMIS in three gripper scaling condition is human-centered transparent when grasping rigid objects. In the *normal* and *quick* gripper scaling conditions, the participants could operate the full range of the remote side gripper without approaching the limit of local manipulator. In these conditions, the system was human-centered transparent, and the grasping kinematics and the gap between action and perception were similar to natural grasping. In the *fine* condition, the relation between the gripper scaling, the size of the gripper, and the size of the objects limited the possible grasping apertures, and the system was not human-centered transparent.

In the action experiment, participants reached and grasped cylinder-shaped objects varying in size. In the *normal* and *quick* gripper scaling conditions, the maximum grasping aperture (MGA) occurred during the reach-to-grasp motion, its mean size was proportional to the size of the object [12], and the variability of MGA violated Weber's law and did not depend on the size of the grasped object. In the perception experiment, participants expressed their perception of the size of the grasped object by opening the remote side gripper to pantomime the perceived size of the object before reaching and grasping cylinder-shaped objects varying in size. The variability of perceptual assessments obeyed Weber's law and increased linearly with the size of the object, consistent with many other psychophysical examples [14], [22]. Similar dissociations between the violation and adherence to Weber's law were reported in many examples of natural grasping [22], including even bimanual grasping [23]. Our results suggest that it is plausible that the grasping in our system was mediated by similar mechanisms in the sensorimotor system. This means that our system is human-centered transparent with respect to grasping of rigid objects.

In contrast, in the *fine* scaling, participants adopted a different behavior in both experiments. The standard deviation of MGA decreased with the object size, and the standard deviation of PS did not depend on the object size. This suggests that with a *fine* gripper scaling, our RAMIS system is not human-centered transparent. We postulate that this violation of human-centered transparency was caused by a ceiling effect on grip aperture. In the *fine* scaling, the minimal necessary aperture for grasping the object was close to the maximum possible grip aperture due to the physical constraint on the at the master side (see dashed black lines and dotted-colored horizontal lines in Fig. 6A and 6B). Therefore, in some trials, the participants hit the limit of the master manipulator's gripper aperture, and in other trials, they tried to avoid hitting it and limited their grip aperture.

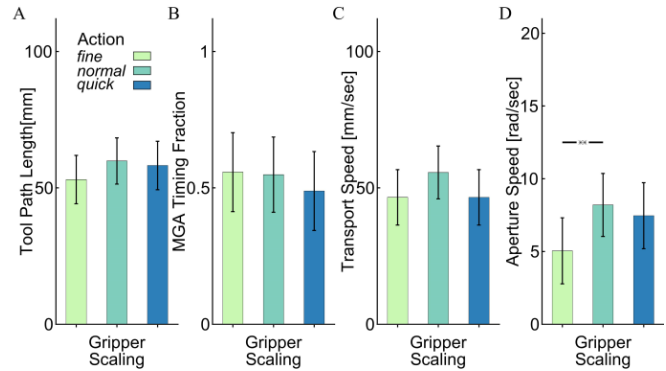

Fig. 9. Kinematic analysis in action trials. Color code as in Fig. 6. (A) Tool path length; (B) maximum grip aperture timing fraction, the MGA timing normalized by total transport time; (C) peak endpoint transport speed; (D) peak grip aperture speed.

In addition, due to the narrow possible gripper aperture range for the larger objects, the participants chose a higher accuracy demand on the grip aperture. This could result in negative dependence of variability of MGA on the size of the object in the action experiment. In the perception experiment, this effect was milder, but nevertheless, could lead to PS variability that did not depend on object's size and violation of Weber's law. Consistently with the speed-accuracy tradeoff [24], the higher accuracy demand also led to a slower grip aperture opening speed in the *fine* gripper scaling (see Fig. 9D). Importantly, this higher accuracy demand was forced by the constraints of the task and the effector, and not by performance accuracy.

An alternative explanation could be that the *fine* gripper scaling enabled participants to be more accurate and reduced the overall variability as well as its dependence on size [25]. This explanation could very well explain the overall reduction in variability, but it is not clear why the larger objects would be affected to a larger degree. This alternative explanation is also not in line with the finding that the variability decreased with object size even for perceptual estimations.

The exact scaling values are specific to our system, task, and objects, and our study is a demonstration of our proposed approach rather than generalizable guideline as to which scaling values are acceptable for human-centered transparency. Currently, the human-centered transparency needs to be evaluated in formal psychophysical experiments for each system and task, and future studies are needed to develop guidelines that may be generalized across systems and tasks.

Interestingly, there are several examples for unnatural grasping that do not involve teleoperation. Grasping of two-dimensional objects that are presented on a computer monitor [26], [27], and performing grasping from memory, or awkward, unpracticed grasping movements also lead to unnatural, perception mediated, grasping [28]. In addition, in a recent study, we showed that teleoperation with transmission delays is not human-centered transparent [29]. These examples show conditions in which grasping is characterized by specific kinematics that indicates that it is also mediated by different neural mechanisms from regular grasping.

In our simple grasping task, the unnatural grasping control in the *fine* gripper scaling did not affect the task performance: all participants successfully lifted the objects without dropping them. Their transport paths were equally straight and fast, and their planning and execution times were similar. Consistent with a classical view of separation between the control of grasping and transport, the choice of the gripper scaling did not affect any of the transport movement parameters, even though this view has been challenged [20], [30], [31]. It may be that if the grasping was followed by an additional task, such as transfer [32] or needle driving [33], we would see performance benefits of our human-centered transparency design. However, future studies are needed to test this hypothesis.

It is important to note that our proposed component of human-centered transparency is not a sufficient condition. Identical kinematics does not necessarily indicate identical underlying neural control mechanisms. Moreover, we only investigated the kinematics of grasping, and ignored other factors, such as the grip force that participants applied on the objects. It is well-documented that in a large variety of motion and force couplings, including the lifting of objects, grip force is modulated in anticipation of the load force [34], [35]. However, when force feedback is not presented to users, they apply a constant grip force during interaction with objects. Adding some form of feedback about the load force of manipulated objects contributes to natural coordination between grip force and load force [36]. Future studies are needed to investigate the effect of adding load and grip force feedback using different bilateral teleoperation architectures on natural grasping.

Finally, in many teleoperation and some RAMIS applications the information transmission may entail delay [37]. Delay in visual feedback may affect the extent of movements [38], [39]. In contrast, when force feedback information is presented with delay, the delay may have dissociable effects on action and perception [40]–[42]. Therefore, a human-centered approach [4] may be used to evaluate and optimize the performance of systems with delayed feedback.

Our human-centered approach is also applicable to other fields of telerobotics research. Transparency and intuitiveness are the main concern for design and control of a telerobotic system for a variety of applications, including working in hazardous environments, such as hot cells and nuclear disaster areas, in inaccessible environments, such as space and underwater [43], or in agricultural robotics. Improved transparency can lead to more efficient and natural, faster to learn and usable systems.

## V. Conclusion

We defined and demonstrated a new human-centered transparency assessment: comparing action and perception in teleoperation grasping of rigid objects to natural grasping. We found that as long as the relation between the gripper scaling, the size of the gripper, and the size of the objects did not limit the possible grasping apertures, our unilateral teleoperated RAMIS system allowed a natural teleoperated grasping with a dissociation between action and perception. Future studies are needed to develop guidelines that may be generalized across systems and tasks, to account for force feedback, and to establish the performance gains of our approach in complicated and clinically relevant surgical tasks.

## VI. Acknowledgment

The authors wish to thank Eli Peretz and Lital Alyagon for their help with the experimental setup and data collection.

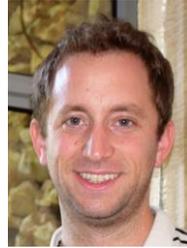

**Amit Milstein** (M'2015) received the M.Sc. degree in Biomedical Engineering from Ben-Gurion University of the Negev, Beer Sheva, Israel, in 2015. He is currently pursuing a PhD in Biomedical Engineering at the Ben-Gurion University of the Negev. His research interests include haptics, human motor control, teleoperation, and robot assisted surgery.

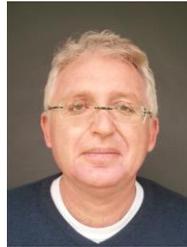

**Tzvi Ganel** Tzvi Ganel received his B.A., M.A. and Ph.D. in psychology from Tel-Aviv University, in 1995, 1998, and 2002, respectively. He is an Associate Professor in the Department of Psychology, Ben-Gurion University of the Negev, where he is the head of the Action-Perception lab. His research interests include object perception, visuomotor control, action-perception dissociations and associations, visual psychophysics, face perception, and selective attention.

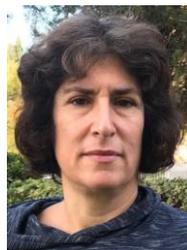

**Sigal Berman** received a B.Sc. in Electrical and Computer Engineering, from the Technion; a M.Sc. in Electrical and Computer Engineering, and a Ph.D. in Industrial Engineering, both from Ben-Gurion University of the Negev in 1991, 1993, and 2003 respectively. She is an associate professor in the Department of Industrial Engineering and Management, Ben-Gurion University of the Negev where she leads the Telerobotics laboratory. Her research interests include robotics, human-robot interaction, learning, and human motor control.

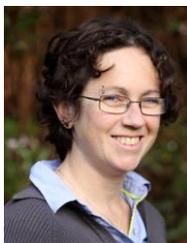

**Ilana Nisky** received the B.Sc (summa cum laude), M.Sc. (summa cum laude), and Ph.D. in Biomedical Engineering from Ben-Gurion University of the Negev, Israel, in 2006, 2009, and 2011, respectively. She is a Senior Lecturer in the Department of Biomedical Engineering, Ben-Gurion University of the Negev, where she is the head of the Biomedical Robotics Lab. Her research interests include human motor control, haptics, robotics, human and machine learning, teleoperation, and robot-assisted surgery.